\newcommand{\ben}{\begin{equation}}
\newcommand{\bal}{\begin{align}}
\newcommand{\een}{\end{equation}}
\newcommand{\eal}{\end{align}}
\newcommand{\bea}{\begin{eqnarray}}
\newcommand{\eea}{\end{eqnarray}}
\newcommand{\nn}{\nonumber\\ }

\newcommand{\qq}{\qquad\qquad}
\newcommand{\QQ}{\qquad\qquad\qquad\qquad}
\newcommand{\al}{{\alpha}}

\newcommand{\infinity}{{\infty}}

\newcommand{\hfb}{{\hfill\break}}
\newcommand{\sub}[1]{_{\stackrel{}{#1}}}
\input amssym.def
\def\Z{{\Bbb Z}}   
   
\documentclass{article}
\usepackage{graphicx}
\usepackage{epsfig}

\begin{document}

\parskip=4pt
\baselineskip=14pt

\title{\vskip-1cm On Robin boundary conditions and the Morse potential in quantum mechanics}
\author{B. Belchev, M.A. Walton\\\\ {\it Department of
Physics and Astronomy,
University of Lethbridge}\\
{\em Lethbridge, Alberta, Canada\ \  T1K 3M4}\\\\
{\small borislav.belchev@uleth.ca, walton@uleth.ca}\\\\
}

\maketitle
\begin{abstract}

The physical origin is investigated of Robin boundary conditions for wave functions at an infinite reflecting wall.  We
consider both Schr\"odinger and phase-space quantum mechanics (a.k.a. deformation quantization), for this simple example of a
contact interaction. A non-relativistic particle moving freely on the half-line is treated as moving on the full line in the
presence of an infinite potential wall, realized as a limit of a Morse potential.  We show that the  wave functions for the
Morse states can become those for a free particle on the half-line with Robin boundary conditions. However, Dirichlet boundary
conditions (standard walls) are obtained unless a mass-dependent fine tuning (to a reflection resonance) is imposed. This
phenomenon was already observed for piece-wise flat potentials, so it is not removed by smoothing. We argue that it explains
why standard quantum walls are standard. Next we consider the Wigner functions (the symbols of both diagonal and off-diagonal
density operator elements) of phase-space quantum mechanics. Taking the (fine-tuned) limit, we show that our Wigner functions
do reduce to the expected ones on the half-line. This confirms that the Wigner transform should indeed be unmodified for this
contact interaction.

\end{abstract}


\vfill\eject

\section{Introduction}

Point interactions and reflecting walls are known as contact interactions \cite{FCT}. In quantum mechanics, they have been a
subject of some interest lately--see \cite{CFT} for some intriguing properties. Perhaps the simplest example is an infinite
reflecting wall \cite{FCT}.

Contact interactions are described by potentials with sharp features. A smooth interaction can be encoded in a potential
feature of a certain width $w$. A contact interaction is obtained for zero width $w$, or, equivalently, infinite sharpness
$\alpha:=1/w$. Physically, these sharp features should be understood in terms of the limit $w\to 0\ (\alpha\to\infty)$.

But can systems with sharp features be quantized after the limit is taken, or is it necessary to quantize before? Do we take
$\al\to\infty$ before or after quantization?\footnote{\, One way to anticipate that it does make a difference is to realize
that the classical limit $\hbar\to 0$ and the sharp limit $\al\to\infty$ do not commute. The wave phenomenon of non-Newtonian
scattering \cite{nNrefl,W} makes that plain. For particle energy exceeding a discontinuous potential,  there is a non-zero
probability  of reflection off the sharp feature, even though the process does not occur classically. Most strikingly, the
probability is independent of Planck's constant, and so does not vanish as $\hbar\to 0$.}

If one takes the sharp limit $\al\to\infty$ before quantization, one can rely on mathematical conditions to proceed. In
operator quantum mechanics,  one only  needs to impose appropriate boundary conditions on wave functions in coordinate space.
The boundary conditions conserve probability  and can be understood as necessary for self-adjointness of Hermitian operators,
like the Hamiltonian, or extensions thereof.\footnote{\, See \cite{SAx} for nice expositions of the theory of self-adjoint
extensions.} For the infinite reflecting wall, Robin boundary conditions are the only possibilities. They include the standard
Dirichlet boundary conditions and the Neumann conditions as two extremal points in a one-parameter continuum of possibilities.

It has been emphasized that the non-standard versions of such interactions should not be ignored, since they may describe
interesting physics \cite{Capri, FCT}.  However, physical considerations such as symmetry (such as time-reversal invariance,
e.g.) can eliminate possibilities in some cases  \cite{CNF}. Can the physical possibilities be restricted in other ways?

Our point of view is that physically, zero-width (or sharp) features {\it must} be understood fundamentally as $\al\to\infty$
limits of nonzero-width (smooth) ones. That is, the sharp case is an idealization, whose treatment should only provide a
shortcut to the results obtained in the physical limit.

In that spirit, the infinite potential wall was described by a limit of the Liouville potential in \cite{KWii}. The standard
wall, with Dirichlet boundary conditions, was recovered. In this work, we will extend the result of \cite{KWii} to the general
case of Robin boundary conditions, by generalizing the Liouville potential to a Morse potential.

In agreement with the results of others \cite{S,FCT}, mass-dependent fine tuning is found to be necessary for non-standard
walls to emerge. We believe that this fine tuning explains why standard quantum walls, with their Dirichlet boundary
conditions, are standard. Non-standard walls are unlikely to be realized physically, because the required fine tuning is
improbable.\footnote{\, This is reminiscent of the result of \cite{CEFGO}, where the renormalization of a different singular
interaction was shown to select a preferred self-adjoint extension.}

Our original motivation came from phase-space quantum mechanics (a.k.a. deformation quantization).\footnote{\, See \cite{ZFC}
for a review and \cite{HWW} for a pedagogical introduction, e.g.} A further complication arises from sharp potential features
in this context \cite{DPi}.\footnote{\, Perhaps this is not surprising, since even the corresponding classical trajectories are
continuous in configuration space but discontinuous in phase-space.} We therefore also examine Wigner functions for
non-standard and standard walls, using the Morse potential.

Dias and Prata \cite{DPi} treated the special (standard) case of Dirichlet boundary conditions for the Schr\"odinger  wave
functions. To describe the complication they found, let $\rho(x,p)$ denote the Wigner function. At finite $\alpha$, it can be
found two ways. First, one can start from the wave functions $\psi(x)$, and build the corresponding density operator. Then a
Wigner transform will yield the Wigner function; denote the result $\rho_\al[\psi](x,p)$. Alternatively, one can use the
dynamical equations of phase-space quantum mechanics. The $\ast$-eigenvalue equations can  be solved, to yield
$\rho_\al[*](x,p)$. As long as $\al<\infty$, we must have \ben \rho_\al[\psi](x,p)\ =\ \rho_\al[*](x,p)\
.\label{finitealWigs}\een

For the case $\alpha=\infty$, Dias and Prata found \ben \rho_\infty[\psi](x,p)\ \not=\ \rho_\infty[*](x,p)\
.\label{infinitealWigs}\een They then {\it assumed} that the Wigner transform $\rho_\infty[\psi](x,p)$ was unaltered and added
a boundary potential so that the $\ast$-eigenvalue equations were compatible. That is, they modified the $\ast$-eigenvalue
equations so that their solutions were $\tilde \rho_\infty[*](x,p) = \rho_\infty[\psi](x,p)$.

In an effort to justify their somewhat ad hoc approach, alternatives to the $\ast$-eigenvalue equations were found in
\cite{KWi}. Dias and Prata then demonstrated \cite{DPii} that the use of the alternative, so-called $\ast$-eigen-$\ast$ value
equations, had a certain equivalence to their treatment. Since the $\ast$-eigen-$\ast$ value equations were derived, rather
than postulated, those arguments provided an indirect justification of their procedure.

More directly, the  limit of the Liouville potential was studied in \cite{KWii}. It was shown there that the Wigner transform
of the wave functions with Dirichlet boundary conditions was indeed physical, as was assumed by Dias and Prata \cite{DPi}. That
is, \ben \lim_{\al\to\infty}\, \rho_\al[\psi](x,p)\ =\ \lim_{\al\to\infty}\, \rho_\al[*](x,p)\ =\ \rho_\infty[\psi](x,p)\
.\label{limrhoalrhoinf}\een The first equality was guaranteed, by (\ref{finitealWigs}), but the second was not.  If the limit
had produced $\rho_\infty[*](x,p)$ instead, for example, then the Wigner transform would have had to be modified, rather than
the $\ast$-eigenvalue equations.

In this work, we will extend the result (\ref{limrhoalrhoinf}) of \cite{KWii} to the general case of Robin boundary conditions.

Let us also mention that in \cite{KWii}, the connection was first made between self-adjoint extensions and the problem
(\ref{infinitealWigs}) found by Dias and Prata \cite{DPi}. Subsequently, those authors were able to show that the Hamiltonians
that included the boundary potentials they introduced were indeed self-adjoint \cite{DPiii}.

This paper is organized as follows. In the next section, we review the realization of Robin boundary conditions in limits of
piece-wise flat potentials, following \cite{{S},{FCT}}. There, mass-dependent fine tuning of the potential was found to be
necessary to realize a non-standard wall, i.e. to avoid the standard Dirichlet boundary conditions. We point out that this fine
tuning is equivalent to selecting a reflection resonance, as defined in \cite{CD}.

In section 3,  the analogous calculation is carried out for a smooth Morse potential. The Robin boundary conditions are
recovered, with the same kind of mass-dependent fine tuning already found in \cite{S,FCT}. We also show that reflection
resonances are again selected in the smooth case.

In section 4, Wigner functions for the Morse potential are considered. Using our solutions of the $\ast$-eigenvalue equations,
described in  \cite{BW}, we show that  the Wigner functions reduce to the expected ones \cite{W} in the appropriate limit. That
is, eqn. (\ref{limrhoalrhoinf}) is indeed obeyed.

The final section is our conclusion.

\section{Robin boundary conditions from a \newline discontinuous potential}

Consider a non-relativistic quantum particle that is confined to the positive half-line with coordinate $x$, but is otherwise
free. Its wave function must satisfy the so-called Robin boundary conditions \ben \psi(0)\ +\ L\,\psi'(0)\ =\ 0\ \ \
\label{Rob} \een for some real length parameter $L\in (-\infty,\infty)\cup\{\infty\}$. The Robin, or mixed boundary conditions
generalise the Dirichlet ($L=0$) and Neumann ($L\rightarrow\pm\infty$) ones. They conserve probability and realize the
self-adjoint extension of the Hermitian Hamiltonian $H = p^2/2m$ on the half-line.

Though there is no mathematical reason other than simplicity to prefer them, Dirichlet boundary conditions are the most
commonly applied. For that reason, infinite walls with other boundary conditions imposed are known as {\it non-standard walls}
\cite{FCT}. In this paper we investigate the physical motivation for standard and non-standard walls.

The real wave function
\ben
 \psi_k(x)\ =\ \sin(k x+\phi)\ \label{sinkxphi}
 \een
obeys the boundary condition (\ref{Rob})
  if the phase is chosen so that
 \ben kL\ =\
-\tan\phi\ .\label{kLtan}
\een
It is appropriate for an unbound particle of energy $\hbar^2k^2/2m$.  For the same dynamics, one bound state also exists, with
(unnormalized) wave function $e^{-x/L}$ and energy $ -\hbar^2/2mL^2$, provided $L>0$.

The bound state provides the length scale $L$: its energy defines it, and its wave function has range $L$. This does not work
for $L<0$, however. A more democratic interpretation is provided by the Wigner time delay (advance) \ben \delta t\ =\ 2\hbar\,
\frac{d\phi}{dE}\ =\ -\, \frac{2mL}{\hbar k(1+k^2L^2)}\ ,\label{wigt}\een for  $L>0$ ($L<0$).\footnote{\, See \cite{FCT} and
references therein.}

Let us now consider a particle moving on the (whole) real line with coordinate $x$ and Hamiltonian \ben H\ =\ p^2/2m\ +\ V(x)\
.\label{HpV} \een A particle with energy $2mE= \hbar^2k^2$ has a time-independent wave function $\psi(x)$ satisfying the
stationary Schr\"odinger equation \ben -\frac{\hbar^2}{2m}\frac{d^2\psi(x)}{dx^2}\ +\ V(x)\,\psi(x)\ =\frac{\hbar^2
k^2}{2m}\,\psi(x)\ .\label{Seqngen} \een We will show that Robin boundary conditions can arise from the limit of a smooth
potential. This generalizes the derivation of  Dirichlet boundary conditions from the sharp $\alpha\to\infty$ limit of the
Liouville potential $V_\alpha(x) =\frac{\hbar^2\kappa^2}{2m}\, e^{-2\alpha x}$. In the context of deformation quantization, the
latter result was obtained in \cite{KWii}.

To prepare for that calculation, we'll first study a discontinuous, piece-wise flat potential: \ben V_\alpha(x)\ =\
\left\{\matrix{ \qq\infinity\ , & \ x < 0\ , \cr -\frac{\hbar^2\kappa^2}{2m}\,\alpha\ell\,(\alpha\ell\,+\, 1 )\ , &\ \ 0\le x
\le 1/\alpha\ ,\cr \qq 0\ , & x> 1/\alpha\ .}\right. \label{SebaV} \een Here $\ell$, $1/\alpha$ and $1/\kappa$ are lengths,
with $\kappa^2>0$ controlling the overall strength of the potential. \v Seba \cite{S} showed that when this potential becomes
an infinite wall as $\alpha\rightarrow\infty$, Robin boundary conditions are recovered.

To see this, solve the Schr\"odinger equation piece-wise to get \ben \psi_\alpha(x)\ =\ \left\{\matrix{ \qq 0\ , & \ x < 0\ ,
\cr \sin\left(x\sqrt{k^2+\kappa^2\,\alpha\ell\,(\alpha\ell\,+\, 1 )} \right) , &\ \ 0\le x \le 1/\alpha\ , \cr A\,\sin( k x +
\phi )\ , & x> 1/\alpha\ }\right. \label{SebaWF}\een for an energy $E=\hbar^2k^2 /2m> 0.$  Notice that the boundary conditions
at $x=0$ are Dirichlet. Those at $x=1/\alpha$, however,  are of the mixed type, i.e., Robin. We can therefore derive Robin
boundary conditions at $x=0_+:= \lim_{\alpha\to\infty} 1/\alpha$. From the point of view of the physical wave function outside
the resulting point interaction, it is the Robin (instead of the Dirichlet) boundary conditions  that must be imposed.

Matching the wave-function values and derivatives at $x=1/\al$, and taking the large $\alpha$ limit gives \ben
 \kappa\ =\ \kappa_n :=\ \frac{\pi}{\ell}\,\left(\, n+\frac 1 2\,\right)\,,\ \ n\in\Z\ .
 \label{discreteka}
 \een
  Then $\sin(\kappa_n\ell) = (-1)^n$, and we find
 \ben
  A\ =\ A_n\ :=\ \sqrt{1+\frac{\pi^4(n+\frac 1 2)^4}{4 k^2 \ell^2}}\ ,
 \ \ \ \ \tan\phi\ =\ \tan\phi_n\ =\ -\frac{2k\ell}{\pi^2(n+\frac 1 2)^2}\ .
 \label{Aphin}
 \een
Comparing to (\ref{kLtan}),
we get
\ben
L\ =\ L_n\ :=\ \frac{2\ell}{\pi^2(n+\frac 1 2)^2}\label{ellLRob}
\een
for the Robin length scale.

So the Robin boundary conditions are found for $x=0_+$, but only barely: there are solutions only for a discrete set of values
of $\kappa$, indexed by the integer $n$. The strength of the potential needs to be finely tuned, tuned differently for
different particle masses,\footnote{\, For the standard wall with Dirichlet boundary conditions, this mass dependence is not
present.} and the non-standard Robin boundary conditions arise for a very limited subset of possible parameters.

What is the physical significance of the fine tuning? It selects a resonance. For this potential the probability of reflection
is always one, but a reflection resonance can still be defined by a rapid change of $\pi$ in the phase shift \cite{CD}. From
the matching conditions  we can derive \ben \frac{\tan (j/\alpha)}{j/\alpha}\ =\ \frac{\tan(k/\alpha + \phi)}{k/\alpha}\ ,
\label{phasedep}\een where $j:= \left[k^2+\kappa^2\alpha\ell(\alpha\ell +1) \right]^{1/2}$. Demanding that
$0=\frac{d^2\phi}{d\phi^2}$, and selecting the maxima of  $\frac{d\phi}{d\phi}$, leads to $\tan(j/\alpha)=\infty$, or \ben
\frac j\alpha\ =\ \frac{\left[k^2+\kappa^2\alpha\ell(\alpha\ell +1) \right]^{1/2}}{\alpha}\ =\ \left( n+\frac 1 2 \right)\,\pi\
,\ \ \ n\in\Z\ \ . \label{refres}\een In the $\alpha\to\infty$ limit, the fine-tuning condition (\ref{discreteka}) is
recovered.

Let us note that the reflection resonance condition (\ref{refres}) corresponds to Neumann boundary conditions at $x=1/\alpha$,
even before the $\alpha\to\infty$ limit is taken. Of course, the requirement (\ref{discreteka}) for Robin boundary conditions
does not select Neumann boundary conditions. Substituting (\ref{discreteka}) yields \ben \frac j\alpha\ =\
 \left( n+\frac 1 2 \right)\,\pi\ +\ (\alpha
L_n)^{-1}\ +\ {\cal O}(\alpha^{-2})\ \   ,  \label{jn}\een using (\ref{ellLRob}). This shows that the fine tuning is to near a
reflection resonance; how it is approached in the $\al\to\infty$ limit determines the Robin length scale $L_n$ and so the
boundary condition that is realized.

Let us also consider the bound states of the \v Seba potential in the $\alpha\to\infty$ limit. For the negative energy states
the wave function will decay exponentially in the interval $x\in (1/\al,\infty)$.  Dividing the matching conditions for the
wave function and its derivative yields \bea \sqrt{-2m\vert E\vert/\hbar^2 +\kappa^2\,\alpha\ell\,(\alpha\ell\,+\, 1 )}
\,\cot\left(\frac{1}{\alpha}\sqrt{-2m\vert E\vert/\hbar^2 +\kappa^2\,\alpha\ell\,(\alpha\ell\,+\, 1 )} \right) \nn
  \ =\ -\sqrt{2m\vert E\vert/\hbar^2}\ .\qq \label{matchbound}
\eea As before we can compare the coefficients in front of the different powers of $\al$. The only possible energy is then
$E=-\hbar^2\kappa^4\ell^2/8m$. Taking into account that (\ref{discreteka}-\ref{ellLRob})  are needed for the  Robin boundary
conditions to arise, we obtain  the  correct energy $-\hbar^2/2mL^2$ in  the $\al\rightarrow\infty$ limit. The bound state
energy is recovered from the \v Seba potential, for $L>0$. We can also verify that, for the same  values of the parameters, the
(unnormalized) wave function of the unique bound state is $e^{-x/L}$ in the limit.

One criticism of these  results could be that Dirichlet boundary conditions were assumed, not derived for the infinite wall
(with no extra structure) at $x=0$ in the \v Seba potential. In addition, infinite potential walls are only idealisations of
very high, but finite walls, and so the infinite wall should be treated as the limit of a finite wall.  However, similar
results were obtained later in \cite{FCT} but with a finite wall, and no particular boundary conditions assumed. Robin boundary
conditions were again obtained, with non-standard walls arising only when a mass-dependent fine tuning was imposed. We will
therefore study here a non-sharp, or smoothed version of the \v Seba potential, rather than of the potential in \cite{FCT}.

The authors of \cite{FCT} speculate that a better choice than their piece-wise flat, discontinuous potential might eliminate
the peculiar mass-dependent fine tuning required for non-standard walls. Presumably, it could also be argued to be possible for
the \v Seba potential \cite{S}. We will find, however, that the mass-dependent fine tuning remains necessary in a smoothed
version of \v Seba's potential. In retrospect, this should perhaps not be surprising, at least for Schr\"odinger quantum
mechanics. The limit that squeezes and stretches the potentials into an infinite wall is so extreme, it seems unimportant
whether the original potential has corners or is smoothed.

\section{Wave functions with Robin boundary\newline conditions from a Morse  potential}

To study how Robin boundary conditions arise as limits in deformation quantization, sharp potential features should be avoided.
We will now carry out an analysis similar to that of the previous section, but for a smooth potential.  The spectrum is first
found for a potential with undetermined parameters. Then we consider a certain limit of the parameters, demanding that we
recover the infinite wall, and that the states realized coincide with the eigenstates for the infinite wall, with Robin
boundary conditions obeyed.

With its short range repulsion and longer range attraction, the smooth Morse potential \bea V(x)\  =\
\frac{\hbar^2\kappa^2}{2m}\,\left(\, e^{-2\alpha x} - b\  e^{-\alpha x} \,\right)\  \label{MorseV} \eea is a rough
approximation to \v Seba's. Besides $\al$, two more parameters are needed --- $\kappa$ determines the overall potential
strength, $b\ge 0$ the position of the well, and together they fix its depth. We will need to impose conditions on these
coefficients in order to obtain Robin boundary conditions in the large $\al$ limit.

The previous section indicates that we need only show that Robin boundary conditions apply at $x=\epsilon$, where epsilon is
very small, but beyond the features of the Morse potential when $\alpha\to\infinity$. For the unbound wave functions,
therefore, we need only require that the relevant wave functions have the asymptotic form $\psi(x)\sim A\sin(k x +\phi)$ as
$x\to\infinity$, with $\phi$ variable.

To do that we first need to find  the unbound wave functions. Following Matsumoto \cite{M}, we can solve the stationary
Schr\"odinger equation for the Morse potential (\ref{MorseV}). The substitution $\psi(x) = \phi(z)$, $z=\exp(-\alpha x)$,
changes the Schr\"odinger equation  into \ben z^2 \phi'' +z \phi'+\frac{1}{\alpha^2}\left[ \frac{2mE}{\hbar^2}-\kappa^2 z^2 +\kappa^2 b\
z\right]\phi=0\ .\label{Morse_schrod} \een This can be further transformed into canonical form (without a first derivative
term) using the substitution $\phi(z)\ =\ z^{-1/2}\,F(z)$. Changing the variables to $y\ :=2\kappa z/ \al$ leads to the
so-called Whittaker equation, treated in \cite{WW}, Chapter XVI: \ben
 f''\ +\ \left\{ -\frac 1 4\ +\ \frac{b\kappa}{2\al}\frac 1 y\ +\ \frac 1{y^2}
 \left[ \frac 1 4 -\left(\frac{i k}{\alpha}\right)^2\right] \right\}\, f\ =\ 0\ , \label{Whittaker}
\een where $f(y):=F(\alpha y/2\kappa)$ and, as before, $k =\sqrt{2 m E}/\hbar$. The two linearly independent solutions are
defined in \cite{Handbook}, p.755. They are called Whittaker functions and can be expressed in terms of the Tricomi confluent
hypergeometric function $U(\mu,\nu,z)$ and the Kummer confluent hypergeometric function $M(\mu,\nu,z)$:\,\footnote{\, The
Whittaker function $M_{lm}(z)$ should not be confused with the Kummer function $M(\mu,\nu,z)$ in the above equation. Subscripts
are used to denote the parameters of the Whittaker functions in the literature, and the explicit bracket notation is used for
confluent hypergeometric functions. For further information involving the hypergeometric functions see \cite{Handbook}, p.753
and \cite{Abram_Stegun}, p.503-506. } \bea
&M_{lm}(z)=z^{m+1/2}e^{-z/2}\,M( 1/2+m-l,1+2m;z)\ ,  \label{Whitt_M} \\
&W_{lm}(z)=z^{m+1/2}e^{-z/2}\,U(1/2 +m-l,1+2m;z)\ .\label{Whitt_W} \eea For our purposes, we only need the definitions of those
functions \bea
M(\mu,\nu;z)=\sum_{n=0}^\infty \frac{(\mu)_n}{(\nu)_n}\frac{y^n}{n!}\ ,\qquad\quad \label{Kummer_M} \\
U(\mu,\nu;z)=\frac{\Gamma(\nu-1)}{\Gamma(\mu)} z^{1-\nu} M(1+\mu-\nu,2-\nu;z)\quad \nn
+\ \frac{\Gamma(1-\nu)}{\Gamma(\mu-\nu+1)}M(\mu,\nu;z)\ .\label{Tricomi}
\eea
Here we use the Pochhammer symbol $(\mu)_n:=\mu(\mu+1)...(\mu+n-1),\ (\mu)_0:=1$.

Now the wave function can be written as \ben \psi_k(x)=e^{\al x/2}\left[C_1 M_{\frac{b\kappa}{2\al},\frac{i k}{\alpha}}(y(x)) +
C_2 W_{\frac{b\kappa}{2\al}\ ,\frac{i k}{\alpha}}(y(x))\right]\,.\label{wave_Fn} \een Imposing reality yields $C_1=0$. The
second term has physical asymptotic behaviour: for large positive $x$ it is sinusoidal with a phase depending on the potential
parameters; for negative $x$ far from the origin, there is the expected rapid exponential decay of a classically forbidden
region. The wave function is therefore \ben \psi_k(x)=C e^{\al x/2} W_{\frac{b\kappa}{2\al},\frac{i k}{\alpha}}
\left(\frac{2\kappa }{\al}e^{-\al x}\right)\ .\label{wave_Fn_final} \een

With the help of equation (\ref{Tricomi}) we can rewrite this  result in a form similar to that given by Matsumoto in \cite{M}
for a Morse potential with $b=2$. The wave function is manifestly real in this form: \bea \psi(y)\ =\ Ce^{-y/2}\,\tilde A \,
y^{i k/\alpha}\, M\left( \frac1 2-\frac{b\kappa}{2\al} + \frac{i k}{\alpha},
1+ \frac{2 i k}{\alpha} ; y \right)+\nonumber\\
+\ Ce^{-y/2}\, \tilde A^*\, y^{-i k/\alpha}\, M \left(\frac1 2-\frac{b\kappa}{2\al} -\frac{i k}{\alpha}, 1- \frac{2 i
k}{\alpha}; y\right)\,, \ \label{MatsumotoWF} \eea with $C$ a real normalization constant, and \bea \tilde
A=\frac{\Gamma(-\frac{2 i k}{\alpha})}{\Gamma\left(\frac1 2- \frac{b\kappa}{2\al}- \frac{ik}{\alpha}\right)}\ .\label{A} \eea

Let us now examine the asymptotic behaviour  of the wave function and how it depends on the parameters. In the limit
$\alpha\to\infinity$, $\exp(-y/2) \sim\exp(-e^{-\alpha x})$ approaches the step function, so the dynamics will be restricted to
the positive half-line. The limit $x\to\infinity$ corresponds to $y\to 0$. Using (\ref{MatsumotoWF}) and (\ref{Kummer_M}) we
obtain \ben
 \psi(x)\ \sim\ C|\tilde A|\,\cos\left[k\, x - \arg(\tilde A) \right]\,. \label{asyi}
\een
 The phase can be calculated from (\ref{A}) and Euler's infinite product formula
\ben \frac 1{\Gamma(u)}\ = \ ue^{\gamma u}\, \prod_{n=1}^\infinity \left[ \left( 1+\frac u n \right) e^{-u/n} \right]\
.\label{Euler} \een  A short calculation shows that
 \bea\label{phase}
\arg(\tilde A) =\frac \pi 2 +\frac{\gamma k}{\alpha}\ - \QQ\QQ\qq \\ \nonumber \sum_{n=0}^\infinity\, \left\{\,  \frac k
{\alpha(n+1)}\  -\ \tan^{-1}\left[\frac{2k}{\alpha(n+1)}\right]\  +\ \tan^{-1}\left[ \frac{2 k}{(2 n+1)\alpha-b\kappa}\right]
\,\right\}\ .\nonumber \eea Apart from the $\pi/2$, all terms will vanish in the $\alpha\to\infty$ limit, except those of the
form $\tan^{-1}\left[ 2 k/\left((2 n+1)\alpha-b\kappa\right)\right]$. For one such term to survive the limit, we need
$\kappa={\cal O}(\al^1)$. If the strength $\kappa$ does  not have this form, we will recover Dirichlet boundary conditions,
i.e. the standard wall. Now, since it is $b\kappa$ that is relevant, we let $b$ absorb the proportionality constant, and use
$\kappa=\al +{\cal O}(\al^0)$. Finally, because the terms of order $\al^0$ and lower will not affect the results, we drop them,
and put  $\kappa=\al$ from now on.

In order to realize Robin boundary conditions (\ref{Rob}), the parameter $b$ must be of the special form
  \ben
b=\ (2 n+1)-\ 2 L^{-1}/\al\ +\ {\cal O}(\al^{-2})\ . \label{sigma_n} \een Here $L$ is a fixed length, independent of $\alpha$.
Then we find
 \ben
  k L  =\tan\arg(\tilde A) \label{etaL}
 \een
in the large $\alpha$ limit, so that the wave function (\ref{MatsumotoWF}) does indeed satisfy the Robin boundary conditions
(\ref{Rob}).

At large $\alpha$ the term $2 L^{-1}/\al$ is negligible compared to the other two. While the parameter $b$ approaches an odd
integer the second infinitesimal term is crucial. Apparently, we need to fine-tune the parameter $b$ to recover the Robin
boundary conditions.  A version of this phenomenon has already been encountered in \cite{FCT} where the parameters can only
take very limited values.  The authors argue that fine tuning may be a result of the particular choice of potential they are
using, possibly because it is not smooth. Since our analysis, using a smooth potential,  produces a version of fine tuning as
well, fine tuning cannot be related to discontinuity alone.

For non-standard walls, we must fine-tune the parameters so that we are near Neumann boundary conditions. Notice that this is
precisely as it was for the \v Seba potential of sect. 2 (see (\ref{jn}) and nearby). Put another way, the fine-tuning is again
to a reflection resonance, or slightly off its peak.

 Let us now consider the bound states. Their wave functions  are given in \cite{FRW}  as
\ben \psi(x)\propto \exp(-\kappa e^{-\al x}/\al)e^{-\al (\nu-b\kappa/2\al +1/2)x} L^{b\kappa/\al-2\nu-1}_{\nu}(2\kappa e^{-\al
x}/\al)\ ,\label{Morse_bound_pre}
\een
 where
  \ben L^\lambda_n(x)=\sum_{m=0}^{n}(-1)^m {{n+\lambda} \choose
{n-m}}\frac{x^m}{m!}\label{Laguerre_assoc} \een
 are the associated Laguerre polynomials $L^\lambda_n(x)$. The energies are
 \ben
 E_\nu=-\frac{\hbar^2\al^2}{2m}(\nu-b\kappa/2\al+1/2)^2\ ,\label{energy_Morse}
 \een
 for integer $\nu\in [0,\lfloor b\kappa/2\al\rfloor]$, where $\lfloor a\rfloor$ is the smallest integer less than $a$.

Consider now the $\al\to\infty$ limit. The Laguerre polynomials  are normalized  to one at zero argument, and $\exp(-\kappa
e^{-\al x}/\al)$ turns into the step function.  The only term that remains to be analyzed  is $e^{-\al (\nu-b\kappa/2\al
+1/2)x}$.  Clearly, $-\al (\nu-b\kappa/2\al +1/2)$ must be a negative constant (independent of $\al$) so that we have a
normalizable wave function that does not disappear in the large $\al$ limit. Again we can set $\kappa=\alpha$, and the solving
for $b$ yields precisely equation (\ref{sigma_n}). Analyzing the bound states provides an alternative way of deriving the fine
tuning condition.

On the other hand, let us give particular values to the constants in (\ref{sigma_n}), i.e. fix  $b$. All the bound states will
vanish for $x>0$ except the one that has highest quantum number $\nu=\lfloor b/2\rfloor$. This is because the maximal integer
will  cancel  the integer part of $b$ and leave only the fine tuning part $-2(L\alpha)^{-1}$. The bound state wave function
$\sim e^{-x/L}$ will be recovered with the correct energy $-\hbar^2/2mL^2$.

To summarize, in this section we demonstrated that the $\al\to\infty$ limit of the Morse potential (\ref{MorseV}) can be used
to generate Robin boundary conditions.  Fine tuning is necessary, however: the parameter $b$ must be an odd non-negative
integer plus a term with $1/\al$ asymptotics  that determine the length scale $L$ of the Robin boundary condition. If the fine
tuning is absent or if the integer part of $b$ is not an odd integer, then we can only recover Dirichlet boundary conditions,
i.e. standard walls. A new observation is that the fine-tuning selects a reflection resonance.

Notice that the definition of $\kappa$ involves the particle mass. So, if the particle mass changes, so must the potential. The
fine tuning required is also mass dependent.

The situation is very similar to that for the discontinuous \v Seba potential \cite{S} treated in sect. 2, and to the results
of \cite{FCT}. The mass-dependent fine-tuning that is necessary for non-standard (Robin boundary condition) walls seems to be
more than an artifact of the choice of potential. In particular, just smoothing out the discontinuities of a piece-wise flat
potential is not sufficient to avoid this property. As already stated, this is perhaps reasonable in hindsight: it seems that
the limit that squeezes and stretches the potentials into an infinite wall is so extreme that it is unimportant whether the
original potential has corners or is smoothed.

\section{Wigner functions and Robin boundary\newline conditions with a Morse potential}

In phase-space quantum mechanics (see \cite{ZFC,HWW}, e.g.), the Wigner function $\rho(x,p;t)$ encodes all measurable
information about the quantum state of a system. It satisfies the equation of motion
 \ben
i\hbar\,\partial_t \rho(x,p,t)=\left [H,\rho(x,p,t)\right ]_\ast\,,\label{dynam_eqn}
\een
 where $\left[H,\rho\right ]_\ast=H\ast\rho -\rho\ast H$, and the Moyal $\ast$-product is defined by \ben \ast\ =\
 \exp\left\{\, \frac{i\hbar}2\, \left(\
\stackrel{\leftarrow}{{\partial}_{x}}\stackrel{\rightarrow}{{\partial}_{p}} -
\stackrel{\leftarrow}{{\partial}_{p}}\stackrel{\rightarrow}{{\partial}_{x}}\, \right) \,\right\}\ \ . \label{Moyal}\een It can
be expressed as a linear combination of stationary Wigner functions with time-dependent coefficients: \ben
\rho(x,p,t)=\sum_{E_L,E_R}C\sub{E_LE_R}e^{-i(E_L-E_R)t/\hbar}\rho\sub {E_L E_R}(x,p)\ .\label{Wigner _function} \een Here
$\rho\sub {E_L E_R}$  denotes the Hamiltonian $\ast$-eigenfunction that can be found by solving the system of equations:
\bea\label{Wigner _function1}
H\ast\rho\sub {E_L E_R}(x,p)=E_L\ \rho\sub {E_L E_R}(x,p)\ ,\\
\rho\sub {E_L E_R}(x,p)\ast H=E_R\ \rho\sub {E_L E_R}(x,p)\ .\label{Wigner _function2} \eea Alternatively, the Wigner transform
\bea\label{Wigner _function3} \rho\sub {E_L E_R}(x,p)=\int_{-\infty}^{\infty}dy\  e^{i y p}\ \langle x+\hbar y/2\vert E_L
\rangle \langle E_R \vert x-\hbar y/2\rangle\  \eea allows them to be determined from the wave functions, if known. For smooth
potentials, the resulting Wigner functions  are known to agree.

For discontinuous potentials, however, that is not necessarily the case \cite{DPi}. For the infinite wall (or a particle
confined to the half-line), Dias and Prata \cite{DPi} showed that the Wigner transform of the density operator only satisfies
the $*$-eigenvalue equations if the free Hamiltonian is modified. No independent motivation was given for the change to the
Hamiltonian, however. It was also assumed that the Wigner transform itself did not need to be adjusted.

An independent motivation was first suggested in \cite{KWii}: self-adjointness of the Hamiltonian. The free Hamiltonian on the
half-line is not self-adjoint. It does have self-adjoint extensions, however, and these correspond precisely to the possible
boundary conditions (\ref{Rob}) (see \cite{SAx}, e.g.).  Subsequently, the self-adjointness of the Dias-Prata modified
Hamiltonian was demonstrated in \cite{DPiii}.

Here we are concerned with the assumption of an unmodified Wigner transform. That is, does the unmodified Wigner transform of
the density operator provide the physical Wigner function? In \cite{KWii}, we answered in the affirmative, by treating the
infinite wall as the limit of a smooth, Liouville potential. Only the standard Dirichlet boundary conditions were recovered,
however. Here we will show that non-standard walls can be realized in a similar way, using the Morse potential, and that the
na\"ive Wigner transform does indeed work, for all Robin boundary conditions,  describing both non-standard and standard walls.

The Wigner transforms of the density operator elements relevant to Robin boundary conditions have already been computed, in
\cite{W}. For $x>0$, using the wave functions (\ref{sinkxphi}, \ref{kLtan}), we find: \bea \rho_\infty[\psi](x,p)\propto
\frac{\sin \left[2 (p/\hbar-k)x\right]}
 {(p/\hbar-k)} +\frac{\sin \left[2
(p/\hbar+k)x\right]}{(p/\hbar+k)}\ \QQ\nn\ +\ 2\cos(2kx-\delta_k)\frac{\sin(  2 x\,p/\hbar)}{p/\hbar}\ .\label{rhoblimit} \eea

In addition, the  $\ast$-eigenvalue equations (\ref{Wigner _function1}, \ref{Wigner _function2}) for the Morse Hamiltonian have
also been solved in \cite{BW}. We must take their limit $\al\to\infty$ as described in the last section, and compare with
(\ref{rhoblimit}).

In \cite{BW}, the $\ast$-eigenvalue equations (\ref{Wigner _function1}, \ref{Wigner _function2}) were solved for the Morse
potential using a Mellin transform and factorization.\footnote{\, The method used there should be useful for any potential that
is a polynomial in an exponential.} Writing \ben E_L\ =:\ \frac{ \hbar^2 k_L^2}{2m}\ ,\ \ \ E_R\ =:\ \frac{ \hbar^2 k_R^2}{2m}\
; \label{ELkLERkR}\een the result was of the form \bea\label{Wigner_function_complete} \rho\sub {E_L E_R}(x,p)\
\propto\QQ\QQ\QQ\nn \qq\int_{c-i\infinity}^{c+i\infinity}ds \left(16e^{4\al x}\right )^{-s}
w_L\left(s-\frac{ip}{2\al\hbar},k_L\right)\, \, w_R\left(s+\frac{ip}{2\al\hbar},k_R\right)\,\ . \eea
 The  factors can be written using
 \bea\label{solution_left}
w_I(t,k_I)\propto \frac{4^{t+ik_I/2\al }\Gamma(-2ik_I/\al)}{\Gamma(1/2-b/2-ik_I/\al )}\Gamma(-2t+ik_I/\al )\times\ \ \ \ \ \ \ \
\ \ \ \ \ \ \ \ \ \ \ \ \ \ \ \ \ \\ \nonumber
_{2}F_{1}\left(1/2-b/2+ik_I/\al ,-2t+ik_I/\al ;1+2\,ik_I/\al ;2\right)+\\ \nonumber
\frac{4^{t-ik_I/2\al }\Gamma(2ik_I/\al )}{\Gamma(1/2-b/2+ik_I/\al)}\Gamma(-2t-ik_I/\al )\times\ \ \ \ \ \ \ \ \ \ \ \ \ \ \ \
\ \ \ \ \ \ \ \ \ \\ \nonumber
_{2}F_{1}\left(1/2-b/2-ik_I/\al ,-2t-ik_I/\al;1-2\,ik_I/\al ;2\right)\,,\,      \eea
 with $I=L,R$. Here we have defined $A=4^{i k_I/\al } \tilde A$, with $\tilde  A$ as in (\ref{A}), and  used the identity
 $\Gamma(2z)\propto 2^{2z-1} \Gamma(z)\Gamma(z+1/2)$. This is the
general solution of the $\ast$-eigenvalue equations of phase-space quantum mechanics, for unbound states in a Morse potential
(\ref{MorseV}) with arbitrary real parameter $b$.

Following \cite{KWii}, we use the residue theorem to find the limit of the Wigner function when $\al\rightarrow\infty$. Since
the calculation is straightforward but lengthy, we omit the details. The integrand of (\ref{Wigner_function_complete}) has $4$
terms, one proportional to $ \tilde  A^2$, one to $ \tilde A^{\ast 2}$ and  two to $\vert \tilde  A\vert^2$. The $\vert \tilde
A\vert^2$-terms yield  contributions proportional to  $[\,e^{2ix(p/\hbar-k)}-e^{-2ix(p/\hbar-k)}]/(p/\hbar-k)$ and
$[\,e^{2ix(p/\hbar+k)}-e^{-2ix(p/\hbar+k)}]/(p/\hbar+k)$. The $\tilde A^2$-term and the $\tilde A^{\ast 2}$-term  yield \bea
\nonumber \hbar[\,e^{2 i \arg \tilde A+2i x p/\hbar- 2 i x k}-e^{2 i \arg \tilde A-2i x p/\hbar- 2 i x k}]/p \eea
 and
\bea\nonumber \hbar[\,e^{-2 i \arg \tilde A+2i x p/\hbar+ 2 i x k}-e^{-2 i \arg \tilde A-2i x p/\hbar+ 2 i x k}]/p\ . \eea Note
that all the terms arising from residues at $i(\pm p/\hbar\pm' k)/2\al+1/2$  produce decaying exponential factors and therefore
do not contribute. Also, the Gauss hypergeometric function $_2F_1(a,b;c;z)$ is analytic with respect to its second argument and
$\tilde \infty$ is its only singularity. The contributions from $_2F_1(a,b;c;z)$ will manifest themselves as a multiplication
by  constants in all cases. In particular, for those terms that survive in the limit of interest, the constant is 1.

The algebra can now be completed to reproduce the Wigner function (\ref{rhoblimit}) for an infinite, but possibly non-standard
wall as we hoped. The Robin boundary conditions are indeed recovered using the Morse potential. We have outlined how the
calculation is done for diagonal elements of the symbol of the density operator, but the non-diagonal case works in similar
fashion.

\section{Conclusion}

Let us  summarize our results.

In section 2, we reviewed \v Seba's analysis \cite{S} showing that Robin boundary conditions (for wave functions) could be
realized by a limit ($\al\to\infty$) of a discontinuous, piece-wise flat potential, eqn. (\ref{SebaV}).  We pointed out that
standard walls (Dirichlet boundary conditions) are generically realized in the sharp limit, and non-standard walls arise only
if a mass-dependent fine-tuning (\ref{discreteka}) is imposed; these observations are in agreement with those made in
\cite{FCT}, for the limit of a similar, but everywhere finite, potential.\footnote{\,  The independent derivation of the
fine-tuning condition from a study of the bound states (see eqn. (\ref{matchbound})), instead of just the continuum, is perhaps
new.} We observe that the parameters are fine-tuned to a reflection resonance in the limit. If the fine-tuning is imposed, then
non-standard walls can be realized, and the Robin length scale $L$ is determined by exactly how the limit resonance is
approached (see eqn. (\ref{jn})).

The analysis of the piece-wise flat \v Seba potential was repeated with a qualitatively similar, but smooth potential, the
Morse potential of eqn. (\ref{MorseV}). Remarkably, the results were almost unchanged. Analysis of both the unbound and bound
states yielded a mass-dependent fine tuning (\ref{sigma_n}) required for non-standard boundary conditions. Again, a reflection
resonance is selected by the fine tuning, and how the resonance is approached in the $\al\to\infty$ limit determines the
precise boundary conditions realized, i.e., the Robin length scale $L$.

As mentioned in the introduction, the infinite reflecting wall is perhaps the simplest example of a so-called contact
interaction. For such, the interaction imposes boundary or matching conditions, such as the Robin boundary conditions on the
half line. Alternatively, the same conditions can be found by demanding that the Hamiltonian or its extension be self-adjoint
(see \cite{SAx}).

Here we have assumed that contact interactions can only be realized physically as limits of smoother, less localized
interactions.\footnote{\, Clearly, we do not consider effective motion for a radial coordinate $r\in[0,\infty)$.} In agreement
with  \cite{S,FCT}, mass-dependent fine tuning was found to be necessary for non-standard walls to emerge. We therefore believe
this explains why standard quantum walls, with their Dirichlet boundary conditions, are standard. Non-standard walls are
unlikely to be realized physically, because the required fine tuning is highly improbable.

It would be interesting to see if the realizations of other contact interactions as limits require similar fine tuning, and if
there are other so-called standard boundary/matching conditions selected that way.

Finally, our primary motivation came from phase-space quantum mechanics, or deformation quantization. In that context, we made
some progress on solving the dynamical equations of Wigner functions, reported in \cite{BW}. Here, we were able to demonstrate
that in the sharp limit, our Wigner functions become those constructed by the Wigner transform from wave functions with Robin
boundary conditions. Specifically, we showed that when $\al\to\infty$, equations (\ref{Wigner_function_complete},
\ref{solution_left}) reduce to the expected Wigner function, eqn. (\ref{rhoblimit}) \cite{W}, $x>0$. This justifies the
assumption that the Wigner transform is unmodified for these examples of contact interactions.

\vskip.25cm \noindent{\bf Acknowledgements}\hfill\break This research was supported in part by a Discovery Grant from the
Natural Sciences and Engineering Research Council of Canada and by the School of Graduate Studies of the University of
Lethbridge. \hfb M.W. thanks the Instituto de Matem\'aticas de UNAM in Morelia, M\'exico, for its warm hospitality.  We also
thank W. Chemissany, S. Das, A. Dasgupta and S. Sur for comments.

\end{document}